\documentclass[3p,times,twocolumn]{elsarticle}
 
\usepackage{lineno,hyperref} 

\journal{Journal of \LaTeX\ Templates} 
\usepackage{multirow}
\usepackage{graphics}

\begin{document}

\begin{frontmatter}

\title{Machine Learning Based  Solutions for Security of Internet of Things (IoT): A Survey}


\author[mymainaddress]{Syeda Manjia Tahsien}

\author[mymainaddress]{Hadis Karimipour\corref{mycorrespondingauthor}}
\cortext[mycorrespondingauthor]{Corresponding author}
\ead{hkarimi@uoguelph.ca}

\author[mymainaddress]{Petros Spachos}

\address[mymainaddress]{School of Engineering, University of Guelph, Guelph, Ontario, N1G 2W1}

\begin{abstract}
 Over the last decade, IoT platforms have been developed into a global giant that grabs every aspect of our daily lives by advancing human life with its unaccountable smart services. Because of easy accessibility and fast-growing demand for smart devices and network, IoT is now facing more security challenges than ever before. There are existing security measures that can be applied to protect IoT. However,  traditional techniques are not as efficient with the advancement booms as well as different attack types and their severeness. Thus, a strong-dynamically enhanced and up to date security system is required for next-generation IoT system. A huge technological advancement has been noticed in Machine Learning (ML) which has opened many possible research windows to address ongoing and future challenges in IoT. In order to detect attacks and identify abnormal behaviors of smart devices and networks, ML is being utilized as a powerful technology to fulfill this purpose. In this survey paper, the architecture of IoT is discussed, following a comprehensive literature review on ML approaches the importance of security of IoT in terms of different types of possible attacks. Moreover,  ML-based potential solutions for IoT security has been presented and future challenges are discussed.
\end{abstract}

\begin{keyword}
Architecture; attack surfaces; challenges; internet of things; IoT attacks; machine learning; security solution.
\end{keyword}

\end{frontmatter}

\section{Introduction}

The  Internet of Things (IoT) interlink electrical devices with a server and exchanges information without any human intervention \cite{ref:1} -\cite{ref:3}. Users can remotely access their devices from anywhere, which makes them vulnerable to different attacks. The security of IoT system is, therefore, a matter of great concern with the increasing number of smart devices nowadays as the devices carry private and valuable information of the clients  \cite{ref:4111} -\cite{ref:4222}. For example, smart home devices and wearable devices hold information about the client's location, contact details, health data, etc. which need to be secured and confidential. Since most of the IoT devices are limited to resources (i.e., battery, bandwidth, memory, and computation), highly configurable and complex algorithm-based security techniques are not applicable \cite{ref:4}.

In order to secure IoT systems, Machine learning (ML) based methods are a promising alternative. ML is one of the advanced artificial intelligence techniques which does not require explicit programming and can outperform in the dynamic networks. ML methods can be used to train the machine to identify various attacks and provide corresponding defensive policy. In this context, the attacks can be detected at an early stage. Moreover, ML techniques seem to be promising in detecting new attacks using learning skills and handle them intelligently. Therefore, ML algorithms can provide potential security protocols for the IoT devices which make them more reliable and accessible than before.

Since only four related comprehensive review articles on ML-based security of IoT have been published until now, there is a need for an up to date literature survey that covers all publications on security of IoT by adopting ML methods. At first, Cui et al. \cite{ref:5} presented a review on different security attacks in IoT and demonstrated various machine learning based solutions, challenges, and research gap using 78 articles till 2017. In 2018, Xiao et al. \cite{ref:6} reviewed different IoT attack models such as spoofing attacks, denial of service attacks, jamming, and eavesdropping and mentioned their possible security solutions based on IoT authentication, access control, malware detections and secure offloading using machine learning techniques. A total of 30 papers have been cited where four different possible ML-based security solutions of IoT devices have been presented. Besides, Chaabouni et al. \cite{ref:8} and more recently, another research group \cite{ref:9} also published a survey paper on machine learning based security of IoT where the authors specifically focused on intrusion detection and various ML related works for the IoT system.

\textbf{Contribution of this Review Paper}\\
Based on the information found so far from literature, the contribution of this paper is as follows:
\begin{itemize}
	\item This literature review concentrates on the ML-based security solutions for IoT systems until most recent articles published in this field till 2019.
	\item At first, an IoT system with its taxonomy of various layers has been presented. Moreover, security in IoT and different potential attacks have been described with their possible layer-wise effects.
	\item This survey will also present different machine learning techniques and their applications to address various IoT attacks.
	\item Besides, a state-of-the-art review has been presented on possible security solutions of IoT devices. It mainly focuses on using different ML algorithms in three architectural layers of the IoT system based on published papers till 2019.
	\item At the end, the authors present possible challenges/limitations in ML-based security of IoT system and their perspective research direction.
\end{itemize}

The rest of the paper is arranged as follows: section II presents an overview of security of IoT consisting of IoT layers and security challenges in IoT ; section III demonstrates attacks in IoT, their effects, and different attack surfaces; section IV describes ML in IoT security including different types of learning algorithms and solutions for IoT security; challenges in ML-based security of IoT has been presented in section V; section VI shows an analysis of published articles on ML-based security of IoT till date; finally, a conclusion of the survey including future recommendations are presented in section VII.
\begin{figure}[t!]
	\centering
	\includegraphics[width= 0.7\linewidth]{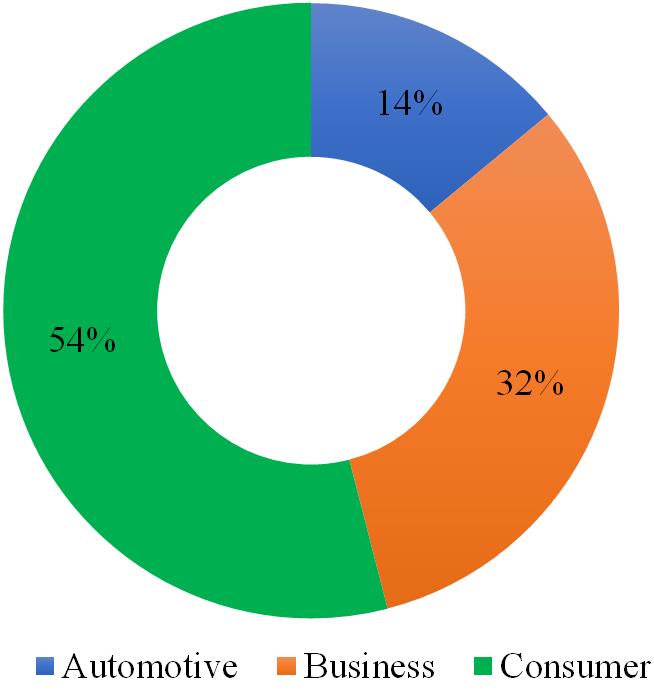}
	\caption{Estimated IoT device users by 2020.}
	\label{fig_p_1}
\end{figure}

\begin{figure}[t!]
	\centering
	\includegraphics[width= 0.8\linewidth]{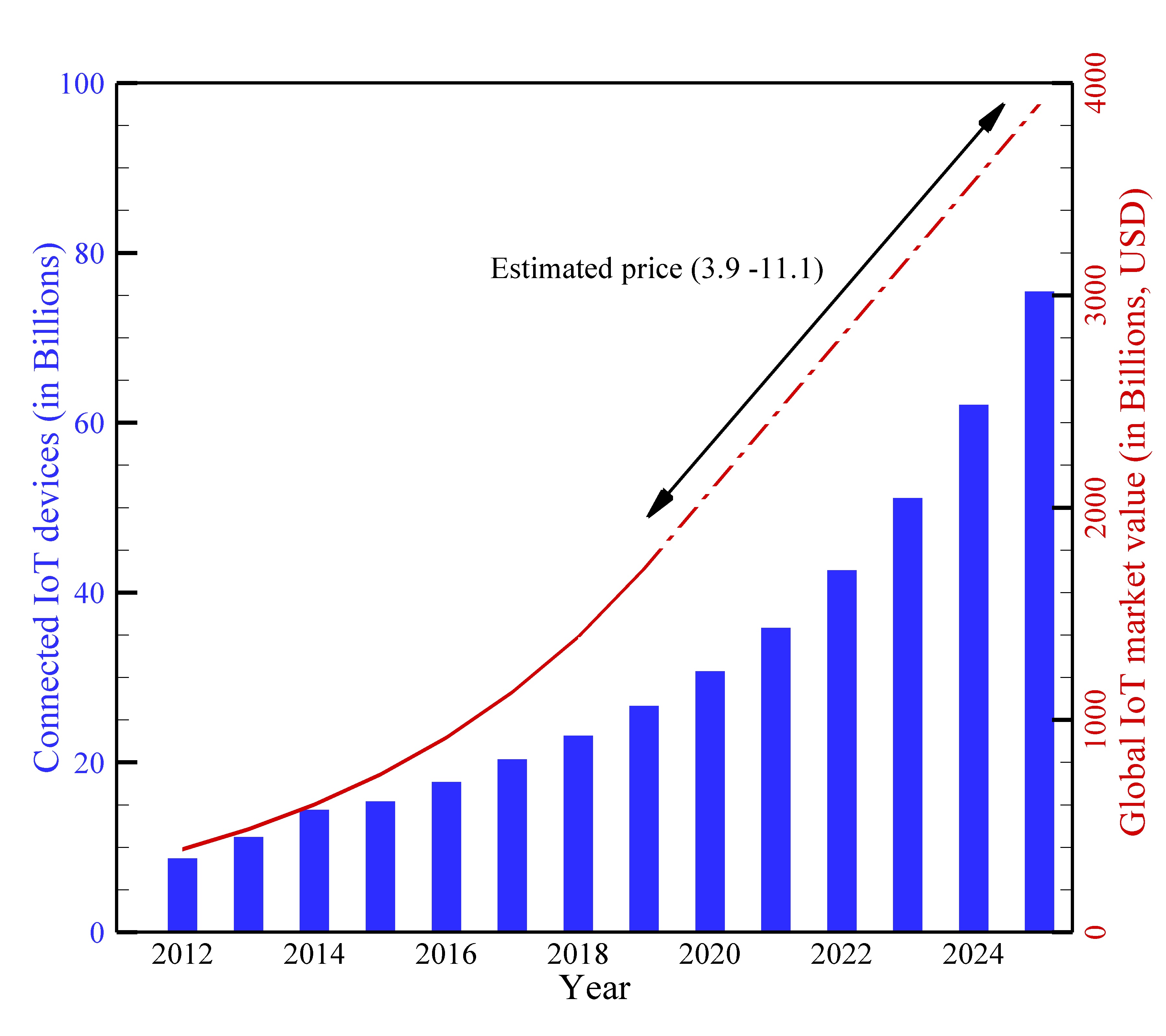}
	\caption{Graphical presentation of total connected IoT devices and global IoT market so far and future prediction.}
	\label{fig_g1}
\end{figure}

\section{Security of Internet of Things}
Security of IoT devices has become a burning question in the twenty-first century. In one side, IoT brings everything close and connects the whole world, on the other hand, it opens various windows to be victimized by different types of attacks.

\begin{figure*}[t!]
	\centering
	\includegraphics[width=0.8\linewidth]{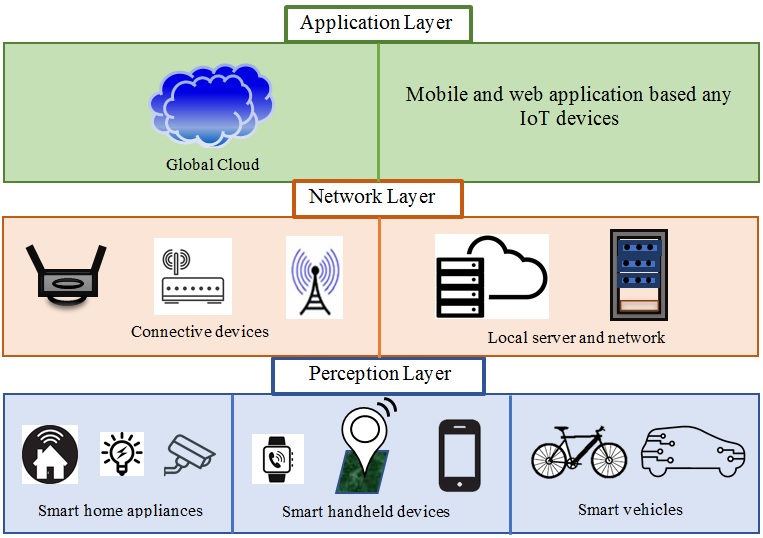}
	\caption{IoT layers architecture.}
	\label{fig_h}
\end{figure*}

Although the term IoT is short in its context wise, it contains the entire world with its smart technologies and services that can be imagined. The word IoT was first used by Kevin Ashton in his research presentation in 1999 \cite{ref:10}. From then, IoT is being used to establish a link between human and virtual world using various smart devices with their services through different communication protocols.

What was a dream 25 years ago is now a reality with the help of IoT. In one word, today's advanced world is wrapped by smart technology and IoT is the heart of it. Now, people cannot think a single moment by themselves without using IoT devices and their services. A survey shows that nearly 50 billion things are going to be connected with internet by 2020 and it will increase exponentially as time passes by \cite{ref:11}. An estimated percentage of IoT device users by 2020 is presented in Fig. \ref{fig_p_1},\cite{ref:11_1}. It is also estimated that IoT is going to capture around $3.9-$11.1 trillion USD economical market by 2025 \cite{ref:12}. The number of connected IoT devices and global market of IoT system so far and future prediction as well until 2025 \cite{ref:12_1,ref:12_2}, is illustrated in Fig.~\ref{fig_g1}.

Therefore, research on IoT and its development and security has received huge attention over the last decades in the field of electrical and computer since. This following two sections will discuss IoT layers and security challenges.

\subsection{IoT Layers}
The architecture of IoT, which is a gateway of various hardware applications, is developed in order to establish a link and to expand IoT services at every doorstep. Different communication protocols, including Bluetooth, WiFi, RFID, narrow and wideband frequency, ZigBee, LPWAN, IEEE 802.15.4, are adopted in different layers of IoT architecture to transmit and receive various information/data \cite{ref:13}, \cite{ref:13_1}.

Moreover, large scale high-tech companies have their own IoT platforms to serve their valuable customers, such as Google Cloud, Samsung Artik Cloud, Microsoft Azure suite, Amazon AWS IoT, etc. \cite{ref:14}. A standard architecture of IoT consists of mainly three layers i.e., perception/physical layer, network layer, and web/application layer \cite{ref:15} as shown in Fig.~\ref{fig_h}.

\subsubsection{Application Layer}
The application layer is the third layer in IoT systems which provides service to the users through mobile and web-based softwares. Based on recent trends and usages of smart things, IoT has numerous applications in this technologically advanced world. Living space/homes/building, transportation, health, education, agriculture, business/trades, energy distribution system, etc. have become smart by the grace of IoT system and it uncounted service \cite{ref:30},\cite{ref:32}.

\subsubsection{Network Layer}
The network layer is more important in IoT systems because it acts as a transmission/redirecting medium for information and data using various connection protocols, including GSM, LTA, WiFi, 3-5G, IPv6, IEEE 802.15.4, etc, which connect devices with smart services \cite{ref:26}. In the network layer, there are local clouds and servers that store and process the information which works as a middle-ware between the network and the next layer \cite{ref:27}-\cite{ref:29}.

Big data is another important factor in the network layer because it attracts the attention of today's ever-growing economical market. The physical objects from the physical layer are producing a huge amount of information/data continuously which are being transmitted, processed, and stored by IoT systems. Since information/data are important for smart services in the network layer, ML and Deep Learning (DL) are extensively used nowadays to analysis the stored information/data to utilize better analysis techniques and extract good uses from it for smart devices \cite{ref:21}.

\subsubsection{Perception Layer}
The first layer of IoT architecture is the perception layer which consists of the physical (PHY) and medium access control (MAC) layers. The PHY layer mainly deals with hardware i.e., sensors and devices that are used to transmit and receive information using different communication protocols e.g., RFID, Zigbee, Bluetooth, etc \cite{ref:16}-\cite{ref:19}.

The MAC layer establishes a link between physical devices and networks to allow to for proper communication. MAC uses different protocols to link with network layers, such as LAN (IEEE 802.11ah), PAN (IEEE 802.15.4e, Z-Wave), cellular network (LTE-M, EC-GSM). Most of the devices in IoT layers are plug and play types from where a huge portion of big data are produced \cite{ref:20}-\cite{ref:25}.

\subsection{Importance of Security in IoT}
 IoT devices are used for various purposes through an open network which makes the devices, therefore, more accessible to the users. In one hand, IoT makes human life technologically advance, easy going, and conformable; on the other hand, IoT puts the users' privacy more in danger due to different threats/attacks \cite{ref:43_1}, \cite{ref:43_2}. Since anyone can access certain IoT devices from anywhere without the user permission, the security of IoT devices has become a burning question. A wide range of security systems must be implemented to protect the IoT devices. However, the physical structure of IoT devices limits its computational functionality which restricts the implementation of complex security protocol \cite{ref:44}. When an intruder accesses a system and exposes private information without the corresponding user's permission, this is considered as a threat/attack \cite{ref:44_1}.

\section{Attacks in IoT}
Over the last few years, the IoT system has been facing different attacks which make the manufacturers and users conscious regarding developing and using IoT devices more carefully. This section describes different kind of attacks, their effects, and attack surfaces in IoT.

\begin{figure}[t!]
	\centering
	\includegraphics[width=\linewidth]{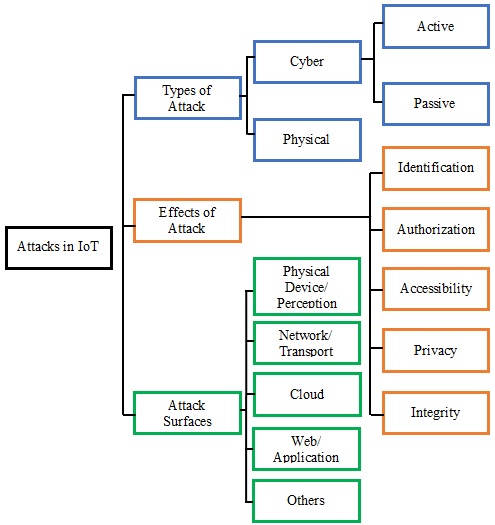}
	\caption{A diagram of a detailed list of IoT security attacks that includes different types of attack, attack surfaces, and attack effects.}
	\label{fig_g}
\end{figure}

\subsection{Types of Attack}
IoT attacks can be classified mainly as cyber and physical attacks where cyber attacks consist of passive and active attacks (see Fig. \ref{fig_g}). Cyber attacks refer to a threat that targets different IoT devices in a wireless network by hacking the system in order to manipulate (i.e., steal, delete, alter, destroy) the user's information. On the other hand, physical attacks refer to the attacks that physically damage IoT devices. Here, the attackers do not need any network to attack the system. Therefore, this kind of attacks are subjected to physical IoT devices e.g., mobile, camera, sensors, routers, etc., by which the attackers interrupt the service \cite{ref:57},\cite{ref:58}.

The following subsections mainly focus on the different types of cyber attacks according to their severeness in IoT devices with Active and passive being the two main categories of a cyber attacks.

\subsubsection{Active Attacks}
\begin{figure*}[t!]
	\centering
	\includegraphics[width=0.9\linewidth]{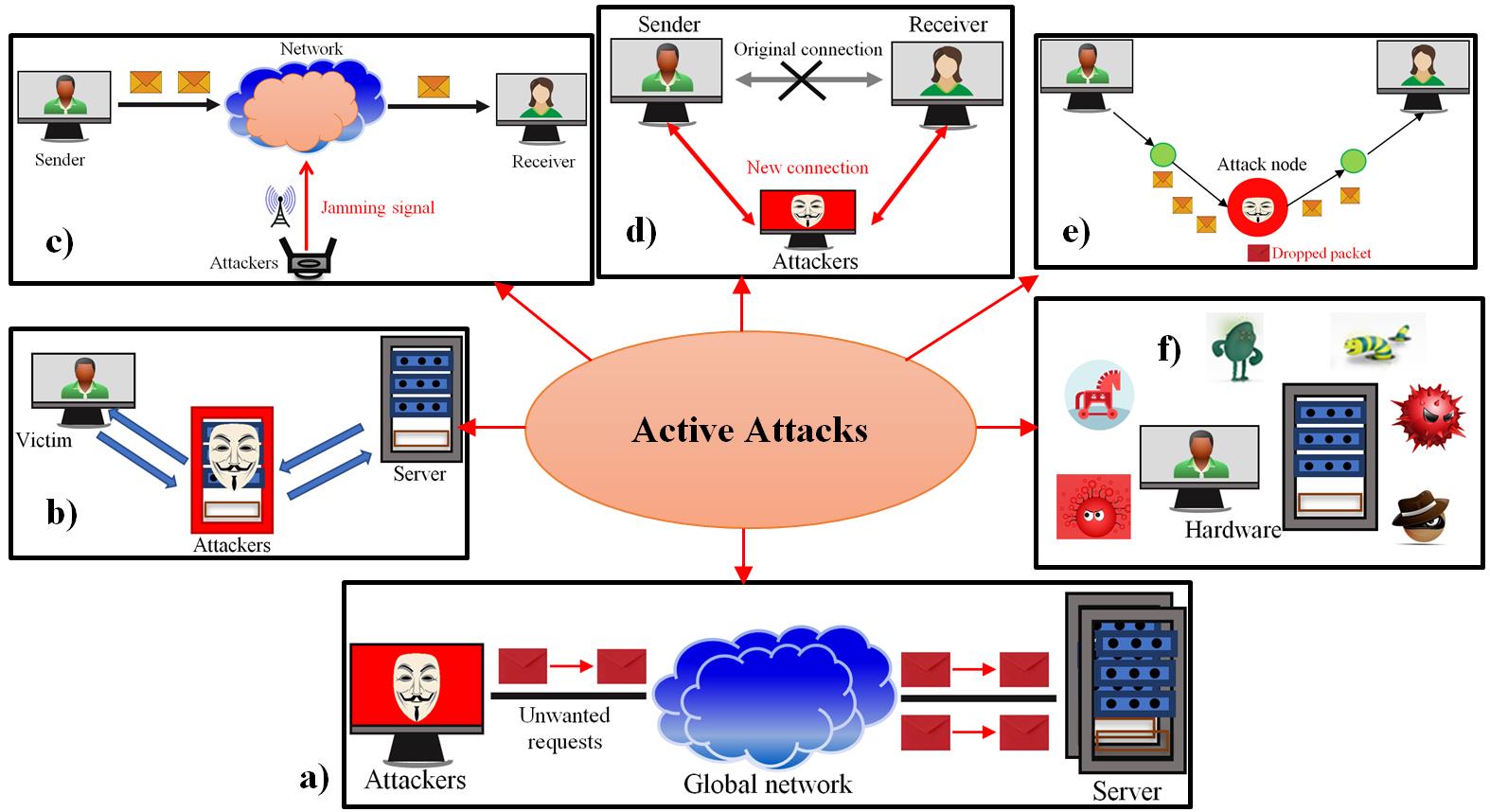}
	\caption{Schematic diagram of different types of cyber attacks: a) Denial of Service attack, b) Spoofing and Sybil attacks, c) Jamming attack, d) Man in the middle attack, e) Selective Forwarding attack, f) Malicious input attack.}
	\label{fig_t1}
\end{figure*}

An active attack happens when an intruder accesses the network and its corresponding information to manipulate the configuration of the system and interrupt certain services. There are different ways to attack IoT device security, including disruption, interventions, and modifications under active attacks. Active attacks such as DoS, man-in-the-middle, sybil attack, spoofing, hole attack, jamming, selective forwarding, malicious inputs, and data tampering, etc. are listed in Table \ref{tab:1} and as illustrated in Fig.~\ref{fig_t1}).

\textit{(i) Denial of Service Attacks}\\
Denial of Service (DoS) attacks are mainly responsible for disrupting the services of system by creating several redundant requests (see Fig. \ref{fig_t1}). Therefore, the user can not access and communicate with the IoT device which makes it difficult to take the right decision. In addition, DoS attacks keep IoT devices always turned on, which can ultimately affect the battery lifetime. A special type of an attack named Distributed DoS (DDoS) attack occurs when consists several attacks happen using different IPs to create numerous requests and keep the server busy. This makes it hard to differentiate between the normal traffic and attack traffic \cite{ref:45}. In recent years, a unique IoT botnet virus named Mirai was responsible for introducing destructive DDoS attacks that have damaged thousands of IoT devices thorugh interferences \cite{ref:46}-\cite{ref:48}.

\textit{(ii) Spoofing and Sybil Attacks}\\
Spoofing and Sybil attacks mainly target the identification (RFID and MAC address) of the users in order to access the system illegally in the IoT system (see Fig. \ref{fig_t1}). It is noticed that TCP/IP suite does not have strong security protocol which makes the IoT devices more vulnerable, especially to spoofing attacks. Moreover, these two attacks initiate further severe attacks, including DoS and man in the middle attacks \cite{ref:49}.

\textit{(iii) Jamming Attacks}\\
Jamming attacks disturb the ongoing communication in a wireless network by sending unwanted signals to the IoT devices which causes problems for the users by keeping the network always busy \cite{ref:50} (see Fig. \ref{fig_t1}). In addition, this attack degrades the performance of the IoT devices by consuming more energy, bandwidth, memory, etc.

\textit{(iv) Man in the Middle Attacks}\\
Man in the middle attackers pretend to be a part of the communication systems where the attackers are directly connected to another user device (see Fig. \ref{fig_t1}). Therefore, it can easily interrupt communications by introducing fake and misleading data in order to manipulate original information \cite{ref:45}.

\textit{(v) Selective Forwarding Attacks}\\
Selective forwarding attack acts as a node in the communication system which allows dropping some packets of information during transmission to create a hole in the network (see Fig. \ref{fig_t1}). This type of attack is hard to identify and avoid.

\textit{(vi) Malicious Input Attacks}\\
Malicious input attacks include malware software attacks, such as trojans, rootkit, worms, adware, and viruses, which are responsible for the damage of IoT devices such as financial loss, power dissipation, degradation of the wireless network performance \cite{ref:4}, \cite{ref:51}, \cite{ref:52} (see Fig. \ref{fig_t1}).

\textit{(vii) Data Tampering}\\
In data tampering, the attackers manipulate the user's information intentionally to disrupt their privacy using unwanted activities. The IoT devices that carry important user's information such as location, fitness, billing price of smart equipment are in great danger to encounter these data tampering attacks \cite{ref:53}.

\subsubsection{Passive Attack}
Passive attacks try to gather the user's information without their consent and exploit this information in order to decrypt their private secured data \cite{ref:54}. Eavesdropping and traffic analysis are the main two ways to perform a passive attack through an IoT network. Eavesdropping mainly deploys the user's IoT device as a sensor to collect and misuse their confidential information and location \cite{ref:55},\cite{ref:56}, \cite{ref:205}.

\subsection{Effects of Attacks}
The effects of IoT attacks are threatening for the network in order to protect the user's privacy, authentication, and authorization. A detailed list of different types of attacks including their effects on IoT devices are presented in Table \ref{tab:1}. The following features need to be considered while developing any security protocol to encounter the attacks for the IoT system.

\begin{table*}[t!]
           
	\centering
	\caption{A list of different kinds of active and passive attacks including their effects.}
	\begin{tabular}{|c|c| p{0.3cm} |}
		\hline
		\textbf{Attack Name} & \textbf{Attack Examples} & \multicolumn{1}{c|}{\textbf{Features: Effected by Attacks}} \\
		\hline
		\multirow{9}[18]{*}\centering{Active} & Sybil Attacks & \multicolumn{1}{c|}{} \\
		\cline{2-2}        & Hole Attacks & \multicolumn{1}{c|}{Identification,} \\
		\cline{2-2}        & Jamming & \multicolumn{1}{c|}{Authorization,} \\
		\cline{2-2}        & Spoofing & \multicolumn{1}{c|}{Accessibility,}\\
		\cline{2-2}        & DoS & \multicolumn{1}{c|}{Confidentiality,} \\
		\cline{2-2}        & Man in the Middle & \multicolumn{1}{c|}{Integrity}\\
		\cline{2-2}        & Selective Forwarding & \multicolumn{1}{c|}{} \\
		\cline{2-2}        & Data tampering & \multicolumn{1}{c|}{} \\
		\cline{2-2}        & Malicious inputs &  \\
		\hline
		\multirow{2}[4]{*}\centering{Passive} & Eavesdropping & \multicolumn{1}{c|}{\multirow{2}[4]{*}\centering{Privacy}}\\
		\cline{2-2}        & Traffic analysis & \multicolumn{1}{c|}{} \\
		\hline
	\end{tabular}%
	\label{tab:1}%
\end{table*}%

\subsubsection{Identification}
Identification refers to the authorization of the user in the IoT network. Clients need to be registered first to communicate with the cloud server. However, trade-offs and robustness of IoT systems create challenges for identification \cite{ref:59}. Sybil and spoofing attacks are responsible for damaging the security of the network and the attackers can easily get access to the server without proper identification. Therefore, an effective identification scheme for the IoT system is necessary which can provide strong security while having system restrictions \cite{ref:60}.

\subsubsection{Authorization}
Authorization deals with the accessibility of the user to an IoT system. It gives permission to only the authorized clients to enter, monitor and use information data of the IoT network. It also executes the commands of those users who have authorization in the system. It is really challenging to maintain all user's logs and give access based on the information, since users are not only confined to humans but also sensors, machines, and services \cite{ref:61}. Moreover, the formation of a strong protective environment is a difficult task while processing the client's large data sets \cite{ref:62}.

\subsubsection{Accessibility}
Accessibility ensures that the services of the IoT system are always rendered to their authorized users. It is one of the important requirements to create an effective IoT network while DoS and jamming attacks disrupt this service by creating unnecessary requests and keep the network busy. Hence, a strong security protocol is needed to maintain the services of IoT devices to be available to their clients without any interruption \cite{ref:63}.

\subsubsection{Privacy}
Privacy is the only factor that both active and passive attacks are facing in IoT system. Nowadays everything, including sensitive and personal information, medical reports, national defense data, etc., are stored and transferred securely through the internet using different IoT devices which are supposed not to be disclosed by any unauthorized users \cite{ref:57}, \cite{ref:65}. However, it is hard to keep most data confidential from unauthorized third parties since attackers can identify the physical location by tracking the IoT device and decrypt the information \cite{ref:32}.

\subsubsection{Integrity}
Integrity property ensures that only authorized users can modify the information of the IoT devices while using a wireless network for communication. This requirement is fundamental for the security of IoT system to protect it from various malicious input attacks such as structured query language (SQL) injection attacks \cite{ref:66}. If this feature is compromised somehow by irregular inspection during data storage in IoT devices, it will affect the functionality of those devices in the long run. In some cases, it can not only reveal the sensitive information but also sacrifice human lives \cite{ref:32}, \cite{ref:67}.

\subsection{Surface Attacks}
\begin{figure*}[t!]
	\centering
	\includegraphics[width=0.75\linewidth]{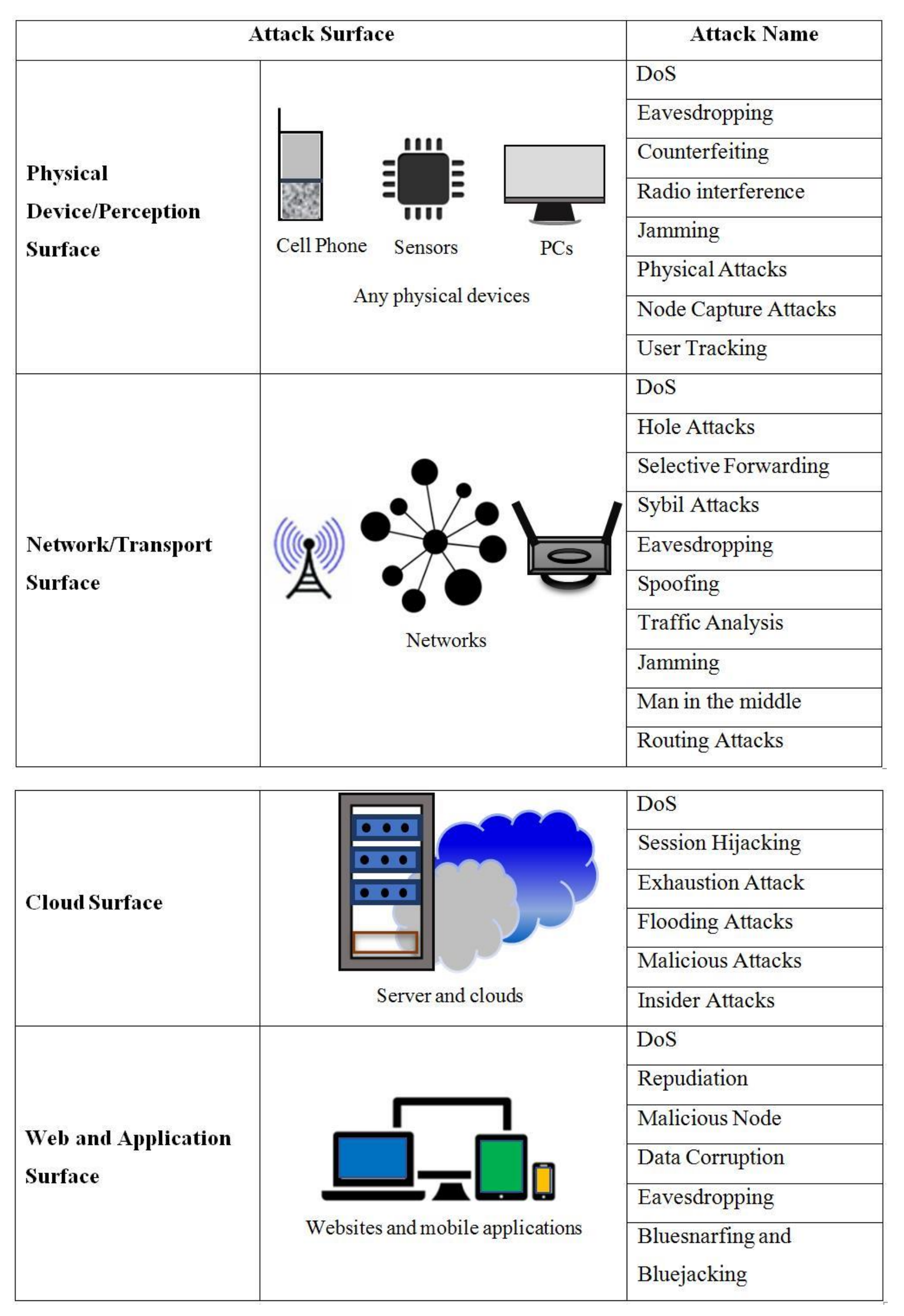}
	\caption{Different attack surfaces of IoT including possible attacks \cite{ref:7}, \cite{ref:69}.}
	\label{fig:i1}
\end{figure*}

The architecture of IoT includes mainly three layers which have been demonstrated in section 2; however, four potential surfaces of IoT have been presented in order to describe attack surfaces more precisely possible attacks besides those three layers in this section (see Fig. \ref{fig:i1}). Here, the IoT surface attacks are categorized as aphysical device/perception surface, network/transport surface, could surface, application/web surface. Moreover, considering the development of smart technologies in IoT system (e.g., smart grid, smart vehicles, smart house, etc.), new surface attacks such as attacks by interdependent, interconnected, and social IoT system are also discussed in this section.

\subsubsection{Physical Device/Perception Surface Attacks}
Physical devices are known as a direct surface attack of the IoT system, since they carry confidential and important information of users. Moreover, attackers can easily access the physical layer of IoT devices. RFID tags, sensors, actuators, micro-controllers, RFID readers are some units of physical devices which are used for identification, communication, collecting and exchanging information, \cite{ref:71}. These parts are vulnerable to DoS, eavesdropping, jamming, radio interference \cite{ref:69}. However, physical attacks are the most alarming for physical device surface.

\subsubsection{Network/Transport Surface Attacks}
Physical devices are connected through network services, including wired and wireless networks in IoT systems. Sensor networks (SNs) play an important role to develop an IoT network. Therefore, wired and wireless sensor network needs to be integrated to construct a large scale IoT surface. This large scale IoT surface is a potential target for different types of attacks as the user's information transfer openly through the sensor networks without any strong security protocol \cite{ref:18}, \cite{ref:69}. In order to launch an attack in a network service surface, attackers will always try to find an open ports or weak routing protocol to access the user network by using their IP address, gateway, and MAC address to manipulate the sensitive information \cite{ref:48}, \cite{ref:74}. Network surface attacks are prone to DoS, jamming, man in the middle, spoofing, Sybil, selective forwarding, traffic analysis, hole attacks, internet attacks, routing attacks, and so on \cite{ref:75}.

\subsubsection{Cloud Surface Attacks}
Besides self-storage capacity, the IoT devices now rely upon the could system which connects most of the smart devices and has unlimited storage capacity \cite{ref:76}. This cloud computing technology enables its stored resources to share remotely for other users \cite{ref:77}, \cite{ref:78}. Cloud computing, therefore, has become the base platform for IoT devices to transport a user's information and store it. Moreover, this could service makes IoT systems dynamic and updates it in a real-time manner \cite{ref:79}-\cite{ref:81}. Therefore, users who are utilizing similar clouds can have their data hacked, stolen, and manipulate through surface attackse. Also, DoS, flooding attacks, insider attacks and malicious attacks can be exposed to cloud surfaces \cite{ref:74}.

\subsubsection{Web and Application Surface Attacks}
Over the last decades, the smart technology is growing very fast which results in increasing demand of IoT devices in order to remote access and control smart devices, such as smart cars, home assistance, watches, glasses, lights and fitness devices. Web and mobile applications make it possible to remotely access and control IoT devices. IoT devices are connected with the network through servers and clouds using a web mobile software based applications. Since there is a technological boom and a merge between the real and virtual world, it is difficult to distinguish between them in the near future. In addition, real-time technology makes IoT devices more alive using smart technologies \cite{ref:82}. Smart devices, such as android operating system based gadgets has attracted the market's attention due to their relatively simple and open architecture and application programming interface \cite{ref:83}, \cite{ref:84}. Therefore, third parties can easily upload their applications on the cloud which creates a way for malware developers to launch different malicious attacks to access IoT devices with/without a user's permission \cite{ref:84}, \cite{ref:85}. Therefore, smart devices that utilize web and mobile applications are vulnerable to DoS, data corruption, eavesdropping, bluejacking, bluesnarfing etc \cite{ref:53}, \cite{ref:87}.

\subsubsection{Other Attacks}
Other new surface attacks are initiated by IoT systems because of a smart technology that is attacked by interdependent, interconnected and social IoT systems \cite{ref:88}, \cite{ref:89}. Attacks that are caused by interdependent IoT systems refer to where the attacker does not need to identify a user's device to attack. For example, a smart building has different kinds of sensors which controls the temperature, air-condition, lighting system. These sensors also depend on other sensors which are connected to the clouds for updating and real-time operation. Since most of IoT devices are interconnected through a global network that creates a wide range of surface attacks for IoT devices, it increases the potential of different types of attacks. Any contaminated treats can easily spread out to other IoT devices because of the interconnected systems. Social surface attacks are new to IoT system due to the increasing number of social sites which involve the user to share their private information with another user. Thus, these social sites may exploit the user's information for any illegal actions \cite{ref:90}, \cite{ref:91}.

\section{Machine Learning (ML) in IoT Security}
ML is one of the artificial intelligence techniques which trains machines using different algorithms and helps devices learn from their experience instead of programming them explicitly \cite{ref:92}. ML does not need human assistance, complicated mathematical equations, and can function in the dynamic networks. In the past few years, ML techniques have been advanced remarkably for IoT security purposes \cite{ref:93}, \cite{ref:94}. Therefore, ML methods can be used to detect various IoT attacks at an early stage by analyzing the behavior of the devices. In addition, appropriate solutions can be provided using different ML algorithms for resource-limited IoT devices. This section is divided into following two subsections i.e., ML Techniques and ML-based solutions for IoT security.

\subsection{ML Techniques}
\begin{figure}[t!]
	\centering
	\includegraphics[width=0.85\linewidth]{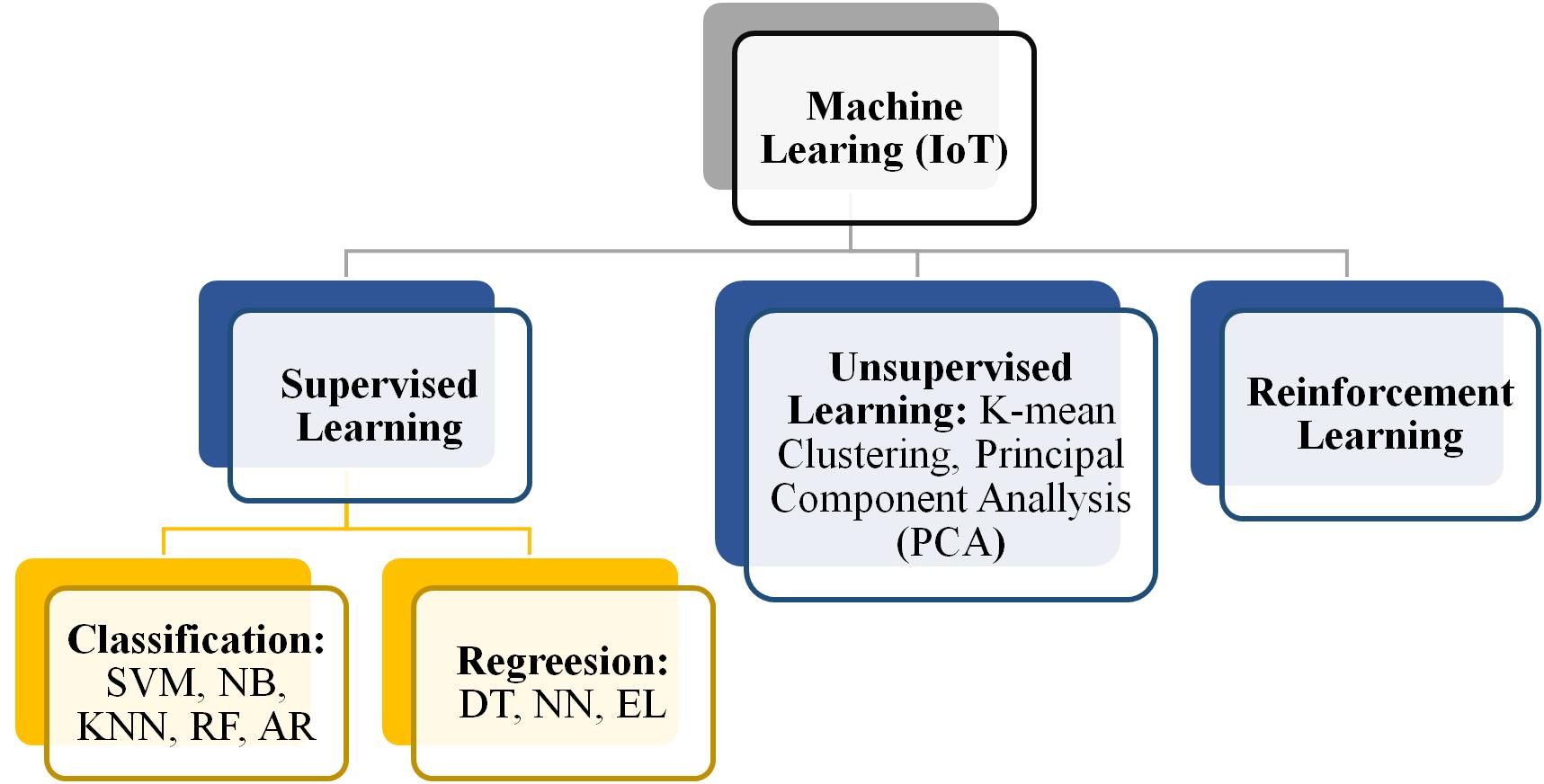}
	\caption{Machine learning and its classification.}
	\label{fig_05}
\end{figure}

ML techniques including supervised techniques, unsupervised techniques, and reinforcement learning can be applied to detect smart attacks in IoT devices and to establish a strong defensive policy. Fig.~\ref{fig_05} illustrates different machine learning algorithms used for the security of the IoT systems.

\subsubsection{Supervised Learning}
Supervised learning is the most common learning method in machine learning where the output is classified based on the input using a trained data set which is a learning algorithm. Supervised learning is classified as classification and regression learning.

\textbf{Classification Learning:}
Classification learning is a supervised ML algorithm where the output is a fixed discrete value/category e.g., [True, False] or [Yes, No], etc. The following subsections will demonstrate different types of classification learning, including Support Vector Machine, Bayesian Theorem, K-Nearest Neighbor, Random Forest, and Association Rule.

\begin{figure}[t!]
	\centering
	\includegraphics[width=0.65\linewidth]{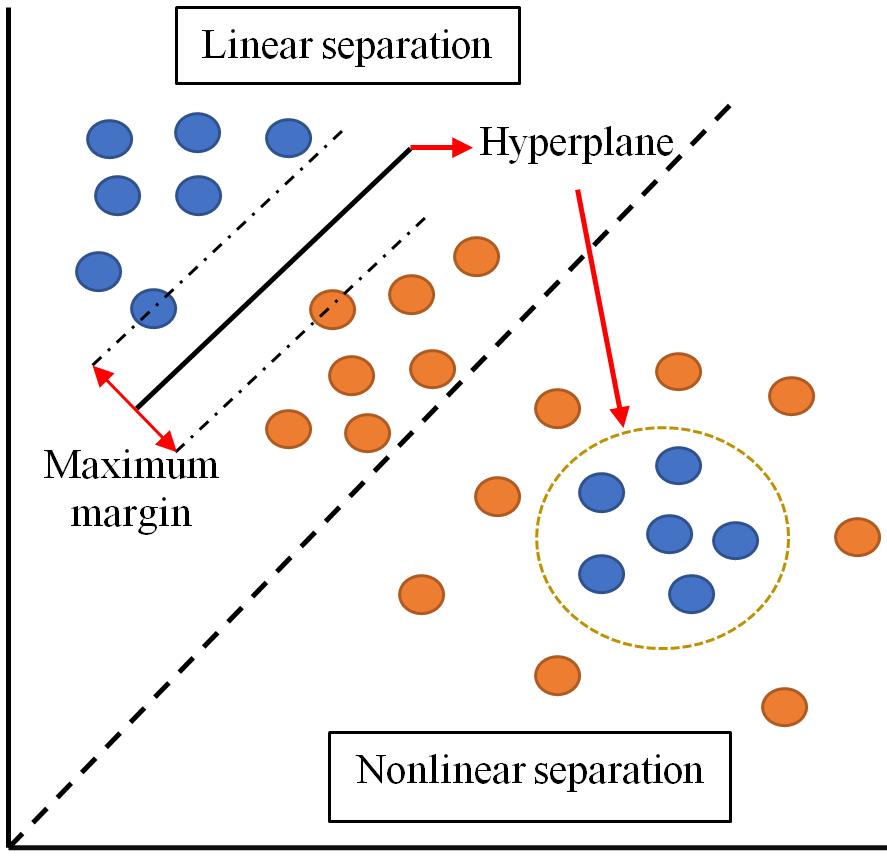}
	\caption{Pictorial illustration of SVM learning techniques to separate the classes for both linear and nonlinear.}
	\label{fig_07}
\end{figure}

\textit{(i) Support Vector Machine (SVM)}\\
SVM algorithm is used to analyze data that use regression and classification analysis. SVM creates a plane named hyperplane between two classes. The goal of the hyperplane is to maximize the distance from each class which distinguishes each class with a minimum error at maximum margin \cite{ref:96}-\cite{ref:97} (as Fig. \ref{fig_07}). If the hyperplane becomes nonlinear after analysis, then SVM uses kernel function to make it linear by adding new features. Sometimes it is hard to use the optimal kernel function in SVM. However, SVM possesses a high accuracy level which makes it suitable for security applications in IoT like intrusion detection \cite{ref:99}-\cite{ref:100}, malware detection \cite{ref:101}, smart grid attacks \cite{ref:102} etc.

\textit{(ii) Bayesian Theorem}\\
The Bayesian theorem is based on the probability of statistics theorem for learning distribution which is known as Bayesian probability. This kind of supervised learning method gets new results based on present information using Bayesian probability. This is known as Nave Bayes (NB). Therefore, NB has been a widely used learning algorithm that needs the prior information in order to implement the Bayesian probability and predict probable outcomes. This is one of the challenges that can successfully be deployed in IoT. NB is usually used in IoT to detect intrusion detection in the network layer \cite{ref:103}, \cite{ref:104} and anomaly detection \cite{ref:105}, \cite{ref:106}. NB has some advantages, such as simple to understand, requiring less data for classifications, easy to implement, applicable for multi-stage calcification. NB depends on features, interactions between features, and prior information which might resist getting accurate outcome \cite{ref:107}.
 
\begin{figure}[!h]
	\centering
	\includegraphics[width=2.2in]{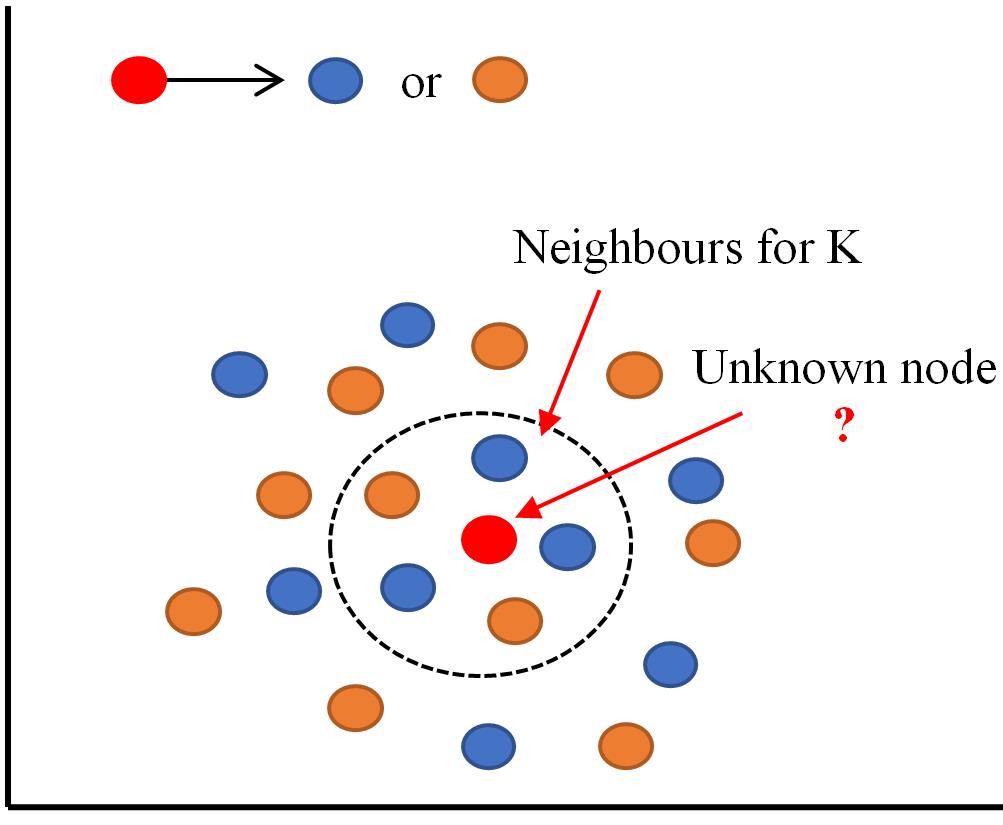}
	\caption{Illustration of KNN learning.}
	\label{fig_08}
\end{figure}

\textit{(iii) K-Nearest Neighbour (KNN)}\\
KNN refers to a statistical nonparametric method in supervised learning which usually uses Euclidian distance \cite{ref:109}. Euclidian distance in KNN determines the average value of unknown node which is k nearest neighbors \cite{ref:111} (see Fig. \ref{fig_08}).  For instance, if any node is lost, then it can be anticipated from the nearest neighbor's average value. This value is not accurate but helps to identify the possible missing node. KNN method is used in intrusion detection, malware detections, and anomaly detection in IoT. KNN algorithm is simple, cheap, and easy to apply \cite{ref:113}-\cite{ref:117}. In contrast, it is a time-consuming process to identify the missing nodes which are challenging in terms of accuracy.

\begin{figure}[t!]
	\centering
	\includegraphics[width=0.55\linewidth]{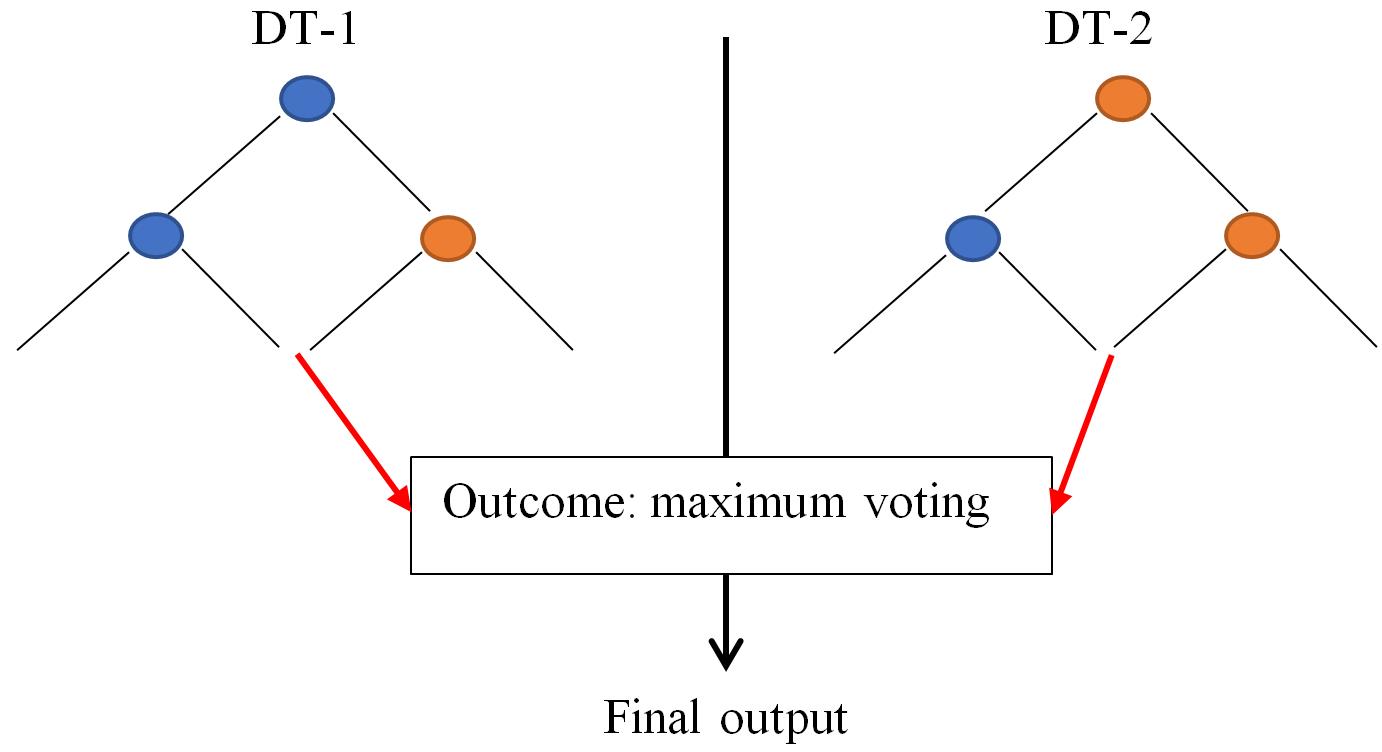}
	\caption{Basic construction of Random Forest learning method.}
	\label{fig_09}
\end{figure}

\textit{(iv) Random Forest (RF)}\\
RF is a special ML method which uses a couple of Decision trees (DTs) in order to create an algorithm to get an accurate and strong estimation model for outcomes. These several trees are randomly developed and trained for a specific action that becomes the ultimate outcome from the model (see Fig. \ref{fig_09}). Although RF uses DTs, the learning algorithm is different because RF considers the average of the output and requires less number of inputs \cite{ref:119}, \cite{ref:120}. RF is typically used in DDoD attack detection \cite{ref:121}, anomaly detection \cite{ref:122}, and unauthorized IoT devices identification \cite{ref:123} in network surface attacks. A previous literature shows that RF gives better result in DDoS attack detection over SVM, ANN, and KNN \cite{ref:121}. Despite RF not being useful in real time applications, it needs a higher amount of training data sets to construct DTs that identify sudden unauthorized intrusions.

\begin{figure}[t!]
	\centering
	\includegraphics[width=0.5\linewidth]{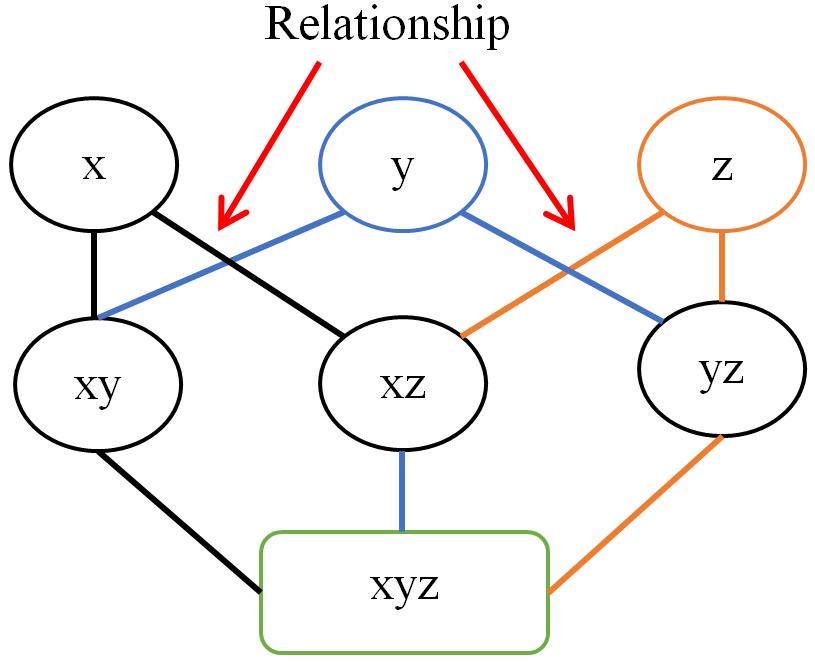}
	\caption{Association rules that establish a relationship between unknown variables.}
	\label{fig_10}
\end{figure}

\textit{(v) Association Rule (AR)}\\
AR method is another kind of supervised ML technique which is used to determine the unknown variable depending on the mutual relationship between them in a given data set \cite{ref:125} (as shown in Fig. \ref{fig_10} ). AR method was successfully used in intrusion detection in \cite{ref:126} where fuzzy AR was used to detect the intrusion in the network. AR is also simple and easy to adopt; however, it is not commonly used in IoT as it has high time complexity and gives results on assumptions that may not provide an accurate outcome for a large and complex model \cite{ref:127}.

\textbf{Regression Learning:}
Regression learning refers to where the output of the learning is a real number or a continuous value depending on the input variables. Different RLs like Decision Tree, Neural Network, Ensemble Learning are presented in the follows subsections.

\begin{figure}[t!]
	\centering
	\includegraphics[width=0.5\linewidth]{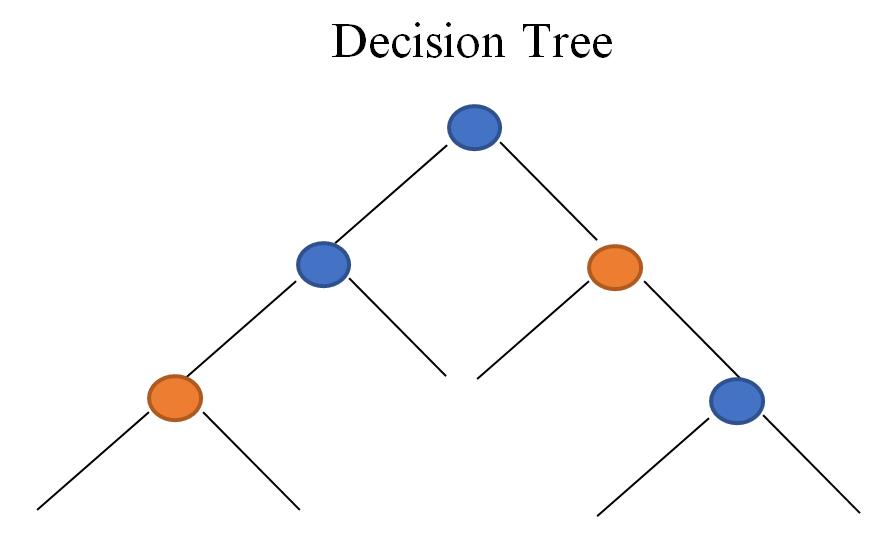}
	\caption{Simple constructor of Decision Tree based learning.}
	\label{fig_11}
\end{figure}

\textit{(i) Decision Tree (DT)}\\
DT is a natural supervised learning method which is like a tree that has branches and leaves. DT has different branches as edges and leaves as nodes (see Fig. \ref{fig_11}). DTs are used to sort out the given samples based on the featured values. DT in ML is mainly categorized as classification and regression \cite{ref:128}. DT has advantages over other ML techniques like simple construction, easy to implement, handling large data samples, and being transparent \cite{ref:129}-\cite{ref:130}. In contrast, this technique has some disadvantages such as requiring a big space to store the data due to its large construction. This makes the learning algorithm more complex if several DTs are considered to eliminate the problem \cite{ref:129}-\cite{ref:130}. DTs are widely used as classifier in security application like DDoS and intrusion detection \cite{ref:132}-\cite{ref:134}.

\begin{figure}[t!]
	\centering
	\includegraphics[width=0.5\linewidth]{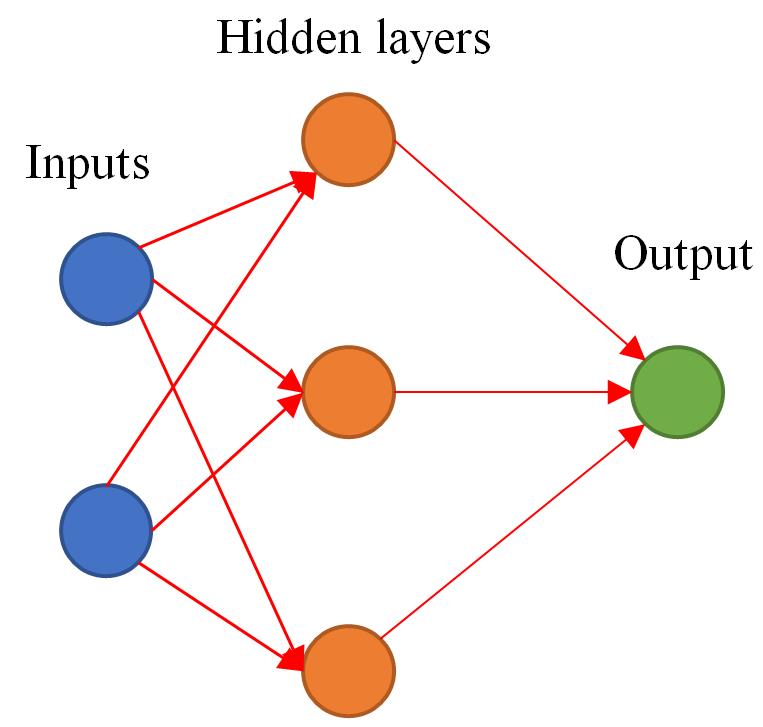}
	\caption{Schematic presentation of Neural Network.}
	\label{fig_12}
\end{figure}

\textit{(ii) Neural Network (NN)}\\
NN technique is constructed based on the human's brain structure which uses neuron. NN has widely used ML techniques that can deal with complex and nonlinear problems \cite{ref:135}-\cite{ref:136}. Hierarchical and interconnected are the two main network categories in NN algorithm based on different functional layers of the neuron (typically: input, hidden and output layers, as shown in Fig. \ref{fig_12}). NN techniques reduce the network response time and subsequently increases the performance of the IoT system. However, NN are computationally complex in nature and hard to implement in a distributed IoT system.

\begin{figure}[t!]
	\centering
	\includegraphics[width=0.45\linewidth]{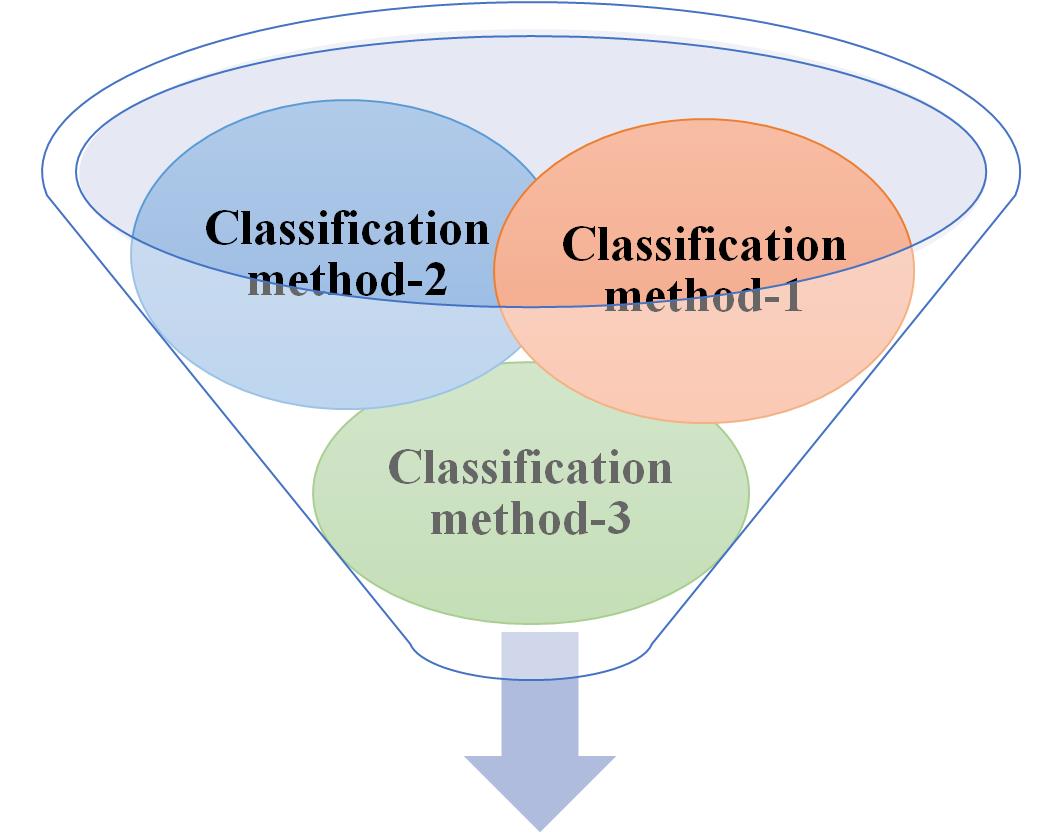}
	\caption{Ensemble learning technique.}
	\label{fig_13}
\end{figure}

\textit{(iii) Ensemble Learning (EL)}\\
EL is a rising learning algorithm in ML where EL uses different classification techniques to get an acceptable outcome by increasing its performance (see Fig. \ref{fig_13}). EL usually combines homogeneous or heterogeneous multi-classifier to get an accurate outcome. Since EL uses several learning algorithms, it is well fitted to solve most problems. However, EL has a high time complexity compared to any other single classifier method. El is commonly used for anomaly detection, malware detection, and intrusion detection \cite{ref:144}-\cite{ref:147}.

\subsubsection{Unsupervised Learning}
In Unsupervised learning, there is no output data for given input variables. Most of the data are unlabeled where the system tries to find out the similarities among this data set. Based on that, it classifies them into different groups as clusters. Many unsupervised learning techniques have been used for security of IoT devices to detect DoS attacks (using multivariate correlation analysis) and privacy protection (applying infinite Gaussian mixture model (IGMM)) \cite{ref:148}, \cite{ref:149}. The following sub-section will focus on the types of unsupervised learning that includes Principal Component Analysis (PCA) and K-means Clustering technique.

\textit{(i) Principal Component Analysis (PCA)}\\
PCA which is also known as a feature reduction technique converts a large data set into smaller ones but holds the same amount of information as in the large set. Therefore, PCA decreases the complexity of a system. This method can be used for selecting a feature to detect real-time intrusion attacks in an IoT system \cite{ref:150}. The combination of PCA and some other ML methods can be applied to provide a strong security protocol. A model proposed by \cite{ref:151} uses PCA and classifier algorithms, such as KNN and softmax regression to provide an efficient system.

\begin{figure}[t!]
	\centering
	\includegraphics[width=0.45\linewidth]{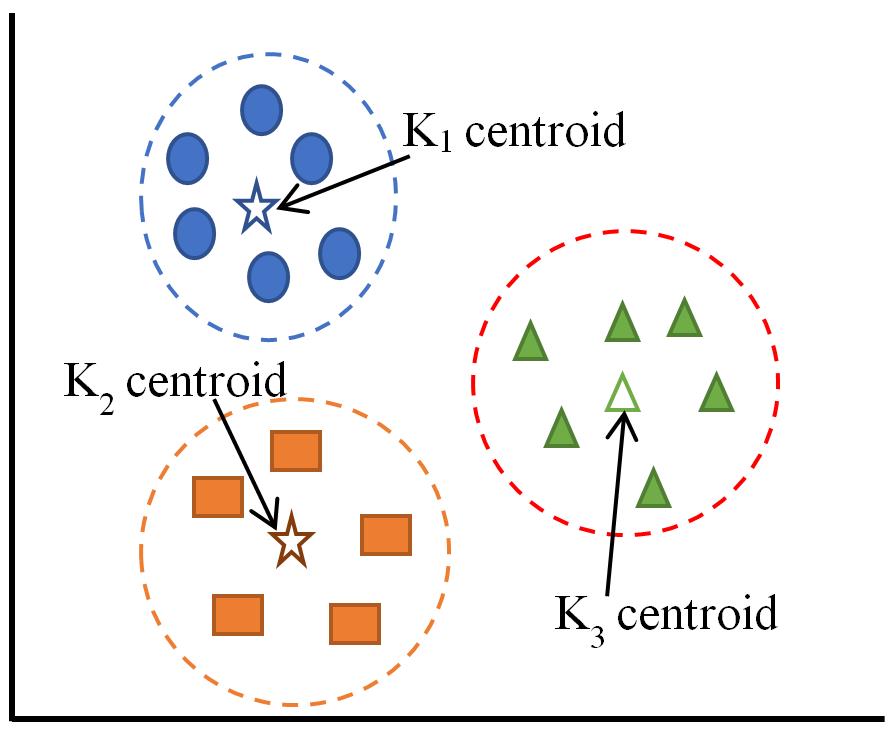}
	\caption{K-mean clustering learning algorithm.}
	\label{fig_14}
\end{figure}

\textit{(ii) K-mean Clustering}\\ 
This unsupervised learning technique creates small groups in order to categorize the given data samples as a cluster. This is a well-known algorithm that uses clustering methods (as shown in Fig. \ref{fig_14}). There are some simple rules to implement this method such as i) Firstly, differentiate the given data set into various clusters where each cluster has a centroid (k-centroid) where the main target is to determine k-centroid for each cluster; ii) Then, select a node from each cluster and relate this with the nearest centroid and keep doing this until every node is contacted. Then, recalculation is performed based on the average value of node from every cluster; iii) Finally, the method redo its  prior steps until it coincides to get the K-mean value \cite{ref:152}-\cite{ref:154}. K-mean learning techniques are useful especially for smart city to find suitable areas for living. K-mean algorithms are also useful in IoT system when labeled data is not required due to its simplicity. However, this unsupervised learning algorithm is less effective compared to supervised learning. K-mean clustering method is usually used in anomaly detection \cite{ref:155}-\cite{ref:157} and Sybil attack detection \cite{ref:158}, \cite{ref:159}.

\subsubsection{Reinforcement Learning (RL)}
RL allows the machine to learn from interactions with its environment (like humans do) by performing actions to maximize the total feedback \cite{ref:160}, \cite{ref:161}. The feedback might be a reward that depends on the output of the given task. In reinforcement learning, there are no predefined actions for any particular task while the machine uses trial and error methods. Through trial and error, the agent can identify and implement the best method from its experience to gain the highest reward.

Many IoT devices (e.g., sensors, electric glass, air conditioner) use reinforcement learning to make changes according to the environment. Moreover, RL techniques have been used for security of IoT devices, including Q-learning, deep Q- network (DQN), post-decision state (PDS), and Dyna-Q to detect various IoT attacks and provide suitable security protocols for the devices. In \cite{ref:49}, \cite{ref:51}, \cite{ref:163}-\cite{ref:166}, Q-learning has been used for authentication, jamming attacks, and malicious inputs whereas Dyna-Q in malware detection and authentication. In addition, DQN and PDS can provide security for jamming attacks and malware detection, respectively \cite{ref:50}.

\subsection{ML based Solution for IoT Security}
ML-based security solutions field for IoT devices has become an emerging research area and is attracting the attention of today's researchers to add more to this field over the last few years. In this section, different ML methods have been presented as a potential solutions for securing IoT systems. These solutions have been investigated based on three main architectural layers of an IoT system, including physical/perception layer, network layer, and web/application layer-wise.

\subsubsection{Physical/Perception Layer}
Traditional authentication methods used for securing the physical surface is not quite sufficient due to the exact threshold value to detect the unwanted signals which give fake alarm \cite{ref:49}. Therefore, ML-based learning methods can be an alternative for authentication in the physical layer. Xiao et al. \cite{ref:49} reported that Q-learning based learning methods reduces the authentication error by about 64.3\% and shows better performance than usual physical layer authentication methods using 12 transmitters. In another study, supervised ML techniques such as Distributed Frank Wolf and Incremental Aggregated Gradient were applied to determine the logistics regression model's parameters in order to reduce the communication overhead and increase the efficiency of spoofing detection \cite{ref:167}. Besides, unsupervised learning like IGMM is also used to secure the physical surface and ensure the authentication of IoT devices \cite{ref:167}.

Research in \cite{ref:168}-\cite{ref:170} showed that RL techniques can effectively address jamming attacks for the security of IoT. A method was proposed for the aggressive jamming attack in \cite{ref:170} where a centralized system scheme was considered. An intelligent power distribution strategy and IoT access point were used to work against the jamming attackers. In another study \cite{ref:50}, RL and deep CNN were combined to avoid jamming signals for cognitive radios that increase RL performance. Cognitive radio (CR) devices have dynamic changing capabilities according to working environments \cite{ref:171}. 

Recently, ML-based a new centralized scheme was proposed in \cite{ref:172} for the security of IoT devices. Basically, it permits certain users with authorization to communicate with the system and safely store authorized users's information. In the proposed peer-to-peer security protocol scheme, clients need to be registered first to the cloud server before starting communication in the IoT system. Besides, Alam et al. \cite{ref:173} proposed a model to avoid attacks and secure IoT devices using Neural Network (NN) and ElGamal algorithm. Here private and public keys were used to control its cryptosystem. Manipulated data have been segmented into groups and then compared with the training data. In addition, a novel defense strategy for detecting and filtering poisonous data collected to train an arbitrary supervised learning model has been presented in \cite{ref:174}.

\subsubsection{Network Layer}
While attack becomes a normal phenomenon, securing network layers becomes a challenge that connects  real life to the virtual world. Accordingly, different supervised ML algorithms like SVM, NN, and K-NN are being used to detect the intrusion attack \cite{ref:97}, \cite{ref:175}-\cite{ref:177}. In one study, NN was used to detect DoS attacks in IoT networks by adopting the multilayer perception based control system \cite{ref:179}.  Saied et al. \cite{ref:180} proposed a model for DDoS attack detection using an ANN algorithm. In the proposed scheme, only real information packets have permission to transmit through the network instead of fake ones. ANN performed better in detecting DDoS attack only if it was trained with updated data sets. Yu et al. \cite{ref:181} research has experimentally showed that SVM based ML method in IoT system was capable of getting a high number of attack detection rate (99.4\%) \cite{ref:181}.

Miettinen et al. \cite{ref:182} presented an IoT SENTINEL model in which the classifier categorizes the IoT devices using RF algorithm to secure it from any unprotected device connection and avoid damage. Meidan et al. \cite{ref:183} used ML classifier algorithms for the identification of IoT devices. Considering various attributes, ML techniques classify the devices according to the connection with the IoT network into two categories (i.e., IoT devices and non-IoT devices). Then, the classifier controls the access of non- IoT devices and prevents possible attacks. A previous study \cite{ref:184} investigated the abnormal behavior of IoT devices and the impact of detection accuracy on ML algorithms (i.e., SVM and k-means) with the partial change of training data sets. A decrement was noticed in accuracy rate for ML techniques and therefore, identification in the variation of accuracy and training data set can be a potential research topic.

An intrusion detection scheme was proposed by \cite{ref:185} at the network layer using ML algorithms for security of IoT devices. Recall, accuracy and precision matrices were used here to evaluate the classifier's performance due to the unbalanced data set. On the other hand, the area under the receiver operating characteristic curve (AUC) can be used as performance matrices for better results \cite{ref:186}, \cite{ref:187}. Along the same direction, ANN techniques were used in \cite{ref:188} to train the machines to detect anomalies in IoT systems. Though the authors found good results from experiments, there is still a scope of further investigations to observe performance with larger data sets in which more data are tampered with attacks. Using unsupervised ML methods, Deng et al. \cite{ref:150}, \cite{ref:189} integrate c-means clustering with PCA and propose an IDS with better detection rate for IoT. In another study, unsupervised ML algorithm (i.e., Optimum-path forest) was also used to develop an intrusion detection framework for the IoT network \cite{ref:192}-\cite{ref:193}.

In 2018, Doshi with his colleagues in \cite{ref:121} presented a way to detect DDoS attacks in local IoT devices using low-cost machine learning algorithms and flow-based and protocol-agnostic traffic data. In this proposed model, some limited behaviors of IoT network such as calculation the endpoints and time taken to travel from one packet to another (time intervals between packets) have been considered. They compared a variety of classifiers for attack detection, including KNN, KDTree algorithm, SVM with the linear kernel (LSVM), DT using Gini impurity scores, RF using Gini impurity scores, NN. It was reported that the proposed techniques can identify DDoS attacks in local IoT devices using home gateway routers and other network middle boxes. The accuracy of the test set for five algorithms is higher than 0.99.

\subsubsection{Web/Application Layer}
K-NN, RF, Q-learning, Dyna-Q- based ML methods have been widely used to secure IoT devices from web/application based attacks, especially for malware detection \cite{ref:51}, \cite{ref:194}. Andrea et al. \cite{ref:194} used supervised ML techniques (both K-NN and RF) to detect malware attacks and reported that RF methods with data set of MalGenome give better detection rate than K-NN. In another research, Q-learning shows better performance in terms of detecting latency and accuracy than Dyna-Q-based detection learning method \cite{ref:51}.

Table \ref{tab:2} presents a list of ML techniques used in different applications to detect attacks as a different layers wise solution of security of IoT.

\begin{table*}[t!]
 \small
	\centering
	\caption{ML based solutions for securing IoT system.}
	\begin{tabular}{|p{1.6cm}|p{6cm}|p{1.6cm}|p{1.50cm}|p{1.4cm}|}
		\hline
		\multirow{2}[2]{*}{\textbf{ML Method}} & \multirow{2}[2]{*}{\textbf{Application/Attack Detection}} & \multirow{2}[2]{*}{\textbf{Layer}} & \multirow{2}[2]{*}{\textbf{Acc. (\%)}} & \multirow{2}[2]{*}{\textbf{Ref.}}\\
		&     &     &     &  \\
		\hline
		\multirow{2}[10]{*}{NN}		  & Security of IoT Networks & \textbf{} & 99  & \cite{ref:194_2} \\
		
		\cline{2-5} & DoS &     &     & \cite{ref:179} \\
		\cline{2-5}        & Intrusion/Malware Detection &     &     & \cite{ref:97}, \cite{ref:194_5} \\
		\cline{2-5}        & Privacy of an IoT Element &     &     & \cite{ref:194_3} \\
		\cline{2-5}        & Security of Mobile Networks &     &     & \cite{ref:194_4} \\
		\hline
		\multirow{2}[8]{*}{KNN} & Intrusion/Malware Detection &     &     & \cite{ref:175},\cite{ref:194} \\
		\cline{2-5}        & Detection of Intrusion, Anomaly, False Data Injection Attacks, Impersonation Attacks & Application, Network  &     & \cite{ref:102},\cite{ref:194_6} \\
		\cline{2-5}        & Authentication of an IoT Element &     & 80  & \cite{ref:194_7} \\
		\hline
		\multirow{3}[22]{*}{SVM} & \multirow{1}[12]{*}{Intrusion/Malware Detection} &     & 97.23 & \cite{ref:194_8},\cite{ref:194_5}\\
		\cline{3-5}        &     &     & 99-99.7 & \cite{ref:194_9},\cite{ref:194_10} \\
		\cline{3-5}        &     &     & 90-92  & \cite{ref:194_15}, \cite{ref:194_13}\\
		\cline{2-5}        & Security of Mobile Networks &     &     & \cite{ref:194_4} \\
		\cline{2-5}        &  False Data Injection Attacks , Authentication, Data Tampering, Abnormal Behaviour& Application, Network, Perception &     &\cite{ref:102}, \cite{ref:169}, \cite{ref:184}, \cite{ref:185} \\
		\hline
		\multirow{2}[4]{*}{DT} & Detection of Intrusion and Suspicious Traffic Sources &     &     & \cite{ref:132} \\
		\cline{2-5}        & Intrusion Detection &     & 50-78 & \cite{ref:194_16} \\
		\hline
		EL  &  Intrusion/Malware Detection , False Data Injection Attacks , Authentication, Data Tampering& Application, Network, Perception &     & \cite{ref:102}, \cite{ref:169}, \cite{ref:184}, \cite{ref:185}\\
		\hline
		\multirow{3}[6]{*}{K-means } & Sybil Detection in Industrial WSNs and Private Data Anonymization in an IoT System, Data Tampering, Abnormal Behaviour & Network  &     & \cite{ref:184} \\
		\cline{2-5}        & Intrusion Detection &     & & \cite{ref:194_17} \\
		\cline{2-5}        & Network attack detection &     & 80.19 & \cite{ref:194_18} \\
		\hline
		\multirow{2}[12]{*}{NB}  &{Intrusion Detection}&     & 50-78 & \cite{ref:194_16},  \cite{ref:194_19}\\
		\cline{2-5}        & Anomaly Detection &     &     & \cite{ref:194_20}\\
		\cline{2-5}        & Security of an IoT Element &     &     & \cite{ref:194_21} \\
		\cline{2-5}        & Traffic Engineering &     & 80-90 & \cite{ref:194_22}\\
		\hline
		\multirow{1}[8]{*}{RF} & \multirow{2}[4]{*}{Intrusion/Malware Detection} &     & 99.67 & \cite{ref:194_23}, \cite{ref:194} \\
		\cline{3-5}        &     &     & 99  & \cite{ref:194_13} \\
		\cline{2-5}        & Anomalies, DDoS, and Unauthorized IoT Devices & Network &     & \cite{ref:182} \\
		\hline
		PCA & Real-Time Detection System, Intrusion Detection & Network &     & \cite{ref:189}\\
		\hline
		\multirow{5}[8]{*}{RL} & DoS &     &     & \cite{ref:166} \\
		\cline{2-5}        & Spoofing &     &     & \cite{ref:49} \\
		\cline{2-5}        & Eavesdropping &     &     & \cite{ref:163} \\ 
                     \cline{2-5} & Jamming &     &     & \cite{ref:50} \\
		\cline{2-5}  & Malware Detection &     &     & \cite{ref:51} \\
		\hline
		AR  & Intrusion Detection &     &     & \cite{ref:126} \\
		\hline

	\end{tabular}%
	\label{tab:2}%
\end{table*}%

\section{Research Challenges}
Currently, the field of IoT and its significance has been reaching at every doorste. Also, the security of IoT has been gaining attention from various networks and application researchers. The application of IoT, its usage, and impact on networks define different challenges and limitations that open new research directions in the future. In order to establish a secured and reliable IoT system, these probable challenges must be addressed. A list of possible challenges and future research fields have been presented based on research that has been conducted so far as well as future predictions in IoT network. In this section, possible research challenges have been presented as follows:

\textit{1) Data Security}:
Any learning algorithm needs a clear and reliable data sample based on what that method can be trained to secure the system. Learning techniques usually observe various attributes of the available data sets and use them to prepare training data sets. In that case, the availability of data, data quality, and data authentication play a vital role to train the data set of the learning methods. Unlike other learning techniques, machine learning also needs large, high quality, and available training data sets to develop an accurate ML technique. If a training data set contains low-quality data which carries noise can interrupt the deploy of a comprehensive and precise learning method. Therefore, authentication of the training data sets is an important challenge in ML techniques for effective security of the IoT network \cite{ref:195}, \cite{ref:196}.

In order to properly implement ML algorithms in IoT system, sufficient data sets are required which are often very difficult to gather based on if the system can identify threats and take necessary actions. In this context, data augmentation is a considerable approach to generate enough data set based on the existing real data. However, the challenge exists where the produced new data samples must properly be distributed in a different class in order to attain maximum accuracy from ML algorithms \cite{ref:195}.

Besides, an exact identification of any attack is another big issue in the security of IoT in order to properly distinguish good from bad state of IoT network.  The challenge is if any intruder knows the attack type and has the ability to manipulate the training data set that is used for ML techniques, then it becomes easy for the attackers to modify their attack types and its effects on the network. Therefore, identifying different kinds of attacks and the probability of their occurrence in the network is a critical future research field in IoT.

\textit{2) Infrastructure Problem}:
When Vender (software-programmer) launches the software, they do not know the weakness of their product which paves a way for the attackers to investigate the infrastructure and hack the system through the software. This type of attack is alarming, and known as zero-day attack, which is very complicated to predetermine with traditional security techniques. Therefore, a strong software infrastructure needs to be developed for the proper security of IoT system. Security must be embedded in every stage in the IoT system starting from hardware to software which will ensure a vulnerable free environment in the overall system. 

\textit{3) Computational Restriction and Exploitation of Algorithms}:
To compile any advanced machine learning algorithm is always challenging because it consumes a large memory and additional energy during processing extensive IoT systems. IoT devices deal with large data sets and with limited resources. Also, if ML methods are incorporated with the IoT system, then they will create more computational complexity for the system. Therefore, there is a need to minimize this complexity using machine learning techniques.

ML methods have been considered for cryptanalysis by attackers which is a potential threat for the IoT system. Though it is usually hard to break the system's cryptography, advanced   ML algorithms, such as SVM and RF are implemented to break strong cryptographic system \cite{ref:197}, \cite{ref:198}.

\textit{4) Privacy Leakage}:
The most common issue in IoT nowadays is privacy. People use smart devices to exchange their data and information for various purposes. Slowly, the information of the clients is being collected and shared which is unknown to the clients. The users are unaware of what, how and where are their private information has been shared. All IoT devices have basic security protocols such as authentication, encryption and security updates.  Therefore, IoT devices require message encryption before sending over the cloud to keep them secret. However, privacy protection must be a security concern in the IoT device design criteria. o illustrate, Google home assistance (Google home speaker and Chromecast have leaked a user's location. Thus, as IoT devices carry confidential and sensitive information/data of the users, there is a possibility for it to be misused if its leaked.

\textit{5) Real-Time Update Issue}:
As IoT devices are increasing rapidly, updating IoT devices' software, firmware update needs to be observed properly. But it is challenging to keep track and apply updates to millions of IoT devices while all devices are not supportive of air update. In that case, applying manual updates is required, such as is real time and data consuming which is cumbersome for users sometimes. Therefore, the term life long learning concept has been introduced to help machines continuously search for updates and makes their firewall strong for updated threats.

Due to the dynamic nature of IoT systems, every day new applications and electronic devices are connected to the network which results in unknown new attacks.  Therefore, this is a challenge of IoT security to adopt an intelligent and real-time updated machine learning algorithm to detect unknown attacks \cite{ref:202}, \cite{ref:203}.

\section{Analysis on Published Articles on ML-based IoT security}
\begin{figure}[t!]
	\centering
	\includegraphics[width=0.9\linewidth]{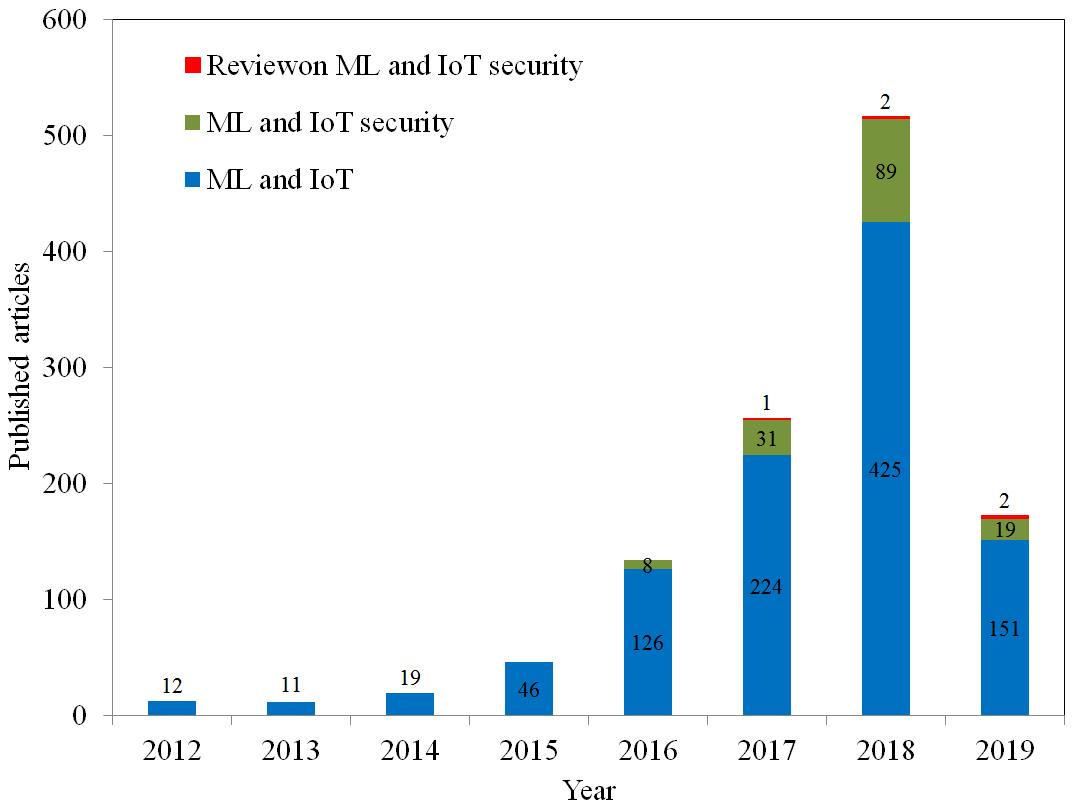}
	\caption{A Statistic on paper published on ML and IoT, ML and security of IoT, and survey on ML and security of IoT till March 2019.}
	\label{fig_p_2}
\end{figure}

Literature shows that ML has been incorporated with IoT since 2002 \cite{ref:204}. Therefore, probable research statistics on ML in IoT, ML in the security of IoT, and review on ML in the security of IoT has been presented in Fig. \ref{fig_p_2} based on the search engine like Elsevier, IEEE, Springer, Wiley, Hindawi, MDPI, Arxiv, and Taylor \& Francis by sorting out to cross-check the title, abstract, and keywords from journal and conference papers. Authors tried their best to incorporate all possible related articles and in this regard, authors manually checked the titles and keywords especially to short out the articles. Fig. \ref{fig_p_2} illustrates that the rate of publication in all cases increases exponentially. Moreover, the publication in ML-based security of IoT starts in 2016 and the growth of publication is very fast which indicates that there is a huge potential of doing research in this field.

\begin{figure}[t!]
	\centering
	\includegraphics[width=0.9\linewidth]{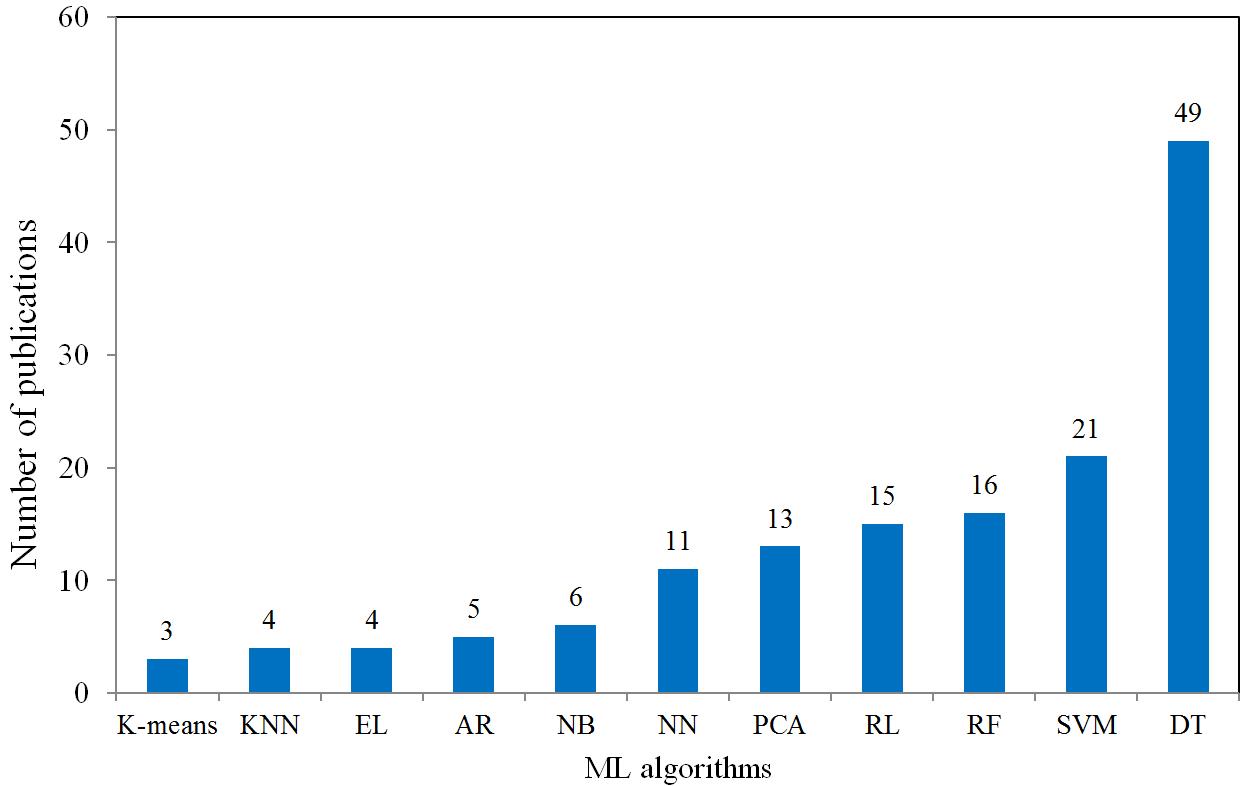}
	\caption{A Statistic on paper published on various ML algorithms used in security of IoT until March 2019.}
	\label{fig_p_3}
\end{figure}

Fig. \ref{fig_p_3} presents statistical results on different ML algorithms based publication in IoT security up to March 2019 which is still increasing with time. It is found that DT was mostly used (32\%) in the security of IoT compared to other learning methods. In addition, these statistics help direct the work of future researchers in potential fields.

\section{Conclusion}
Internet of Things (IoT) have the ability to change the future and bring global things into our hand. As a result, anyone can access, connect, and store their information in the network from anywhere using the blessing of smart services of IoT. Although, the empowerment of IoT connects our lives with the virtual world through smart devices to make life easy, comfortable, and smooth, security becomes a great concern in IoT system to care for its services. Therefore, to enhance the security with time and growing popularity, challenges and security of IoT has become a promising research in this field which must be addressed with novel solutions and exciting strategic plans for uncertain attacks in upcoming years. In this paper, a state of the art comprehensive literature review has been presented on ML-based security of IoT that includes IoT and its architecture, a thorough study on different types of security attacks, attack surfaces with effects, various categories of ML-based algorithms, and ML-based security solutions. In addition, research challenges have been demonstrated. Comparing with other review papers, this literature survey includes all papers on IoT and ML-based security of IoT up to 2019. During 2018, there was a huge acceleration in research on security of IoT. This literature review has focused on ML embedded algorithms on security of IoT from where anyone can get a general idea about different potential IoT attacks and their surface wise effects. Also, ML algorithms have been discussed with possible challenges that can aid future researchers to fix their ultimate goals and fulfill their aim in this field.

\section*{References}


\end{document}